\documentclass[notitlepage,superscriptaddress,showpacs,nobalancelastpage,twocolumn,aps,prb]{revtex4-1}
\usepackage[english]{babel}
\RequirePackage[T1]{fontenc}
\RequirePackage{times} 

\usepackage{amsfonts}
\usepackage{subfigure}
\usepackage{amsmath}
\usepackage{amssymb}
\usepackage{bm}
\usepackage{graphicx,epstopdf}
\usepackage{appendix}
\usepackage{array}
\usepackage{color}
\usepackage[hidelinks]{hyperref}
\usepackage{cancel}
\usepackage{my_symbols}
\usepackage[normalem]{ulem}
\usepackage{CJK}
\providecommand{\U}[1]{\protect\rule{.1in}{.1in}}

{\par}

\newcommand\rmv{\bgroup\markoverwith {\textcolor{red}{\rule[0.5ex]{2pt}{0.4pt}}}\ULon}

\begin{document}
\begin{CJK*}{UTF8}{gbsn} 
\title{Cavity-mediated dissipative spin-spin coupling}
\author{Vahram L. Grigoryan}
\affiliation{Institute for Quantum Science and Engineering, Southern University of Science and Technology, Shenzhen 518055, China}
\author{Ke Xia}
\email[Corresponding author:~]{kexia@bnu.edu.cn}
\affiliation{Institute for Quantum Science and Engineering, Southern University of Science and Technology, Shenzhen 518055, China}

\begin{abstract}
We study dissipative spin-spin coupling in dispersive regime mediated by virtual photons in a microwave cavity. Dissipative coupling between magnetization of each magnetic material and the cavity photons is established by means of two phase shifted driving forces acting on each magnetization. We show that when only one of the magnetization is dissipatively coupled to the cavity, the cavity-mediated spin-spin coupling too, exhibits mode level attraction in the spectrum. By tuning the phase parameter at each ferromagnetic insulator we can shift the order of "dark" and "bright" collective modes with phase difference equal to $0$ or $\pi$. Moreover, by selectively applying the phase shifted field it is possible to construct "dark" and "bright" collective modes with phase difference equal to $\pm \pi/2.$
\end{abstract}
	\maketitle
\end{CJK*}

\section{Introduction}
Recent progress in hybridization of magnons (collective spin excitations) in yttrium iron garnet (YIG) ferrimagnetic insulator (FI) with microwave cavities makes the coupled magnon-photon system a good candidate for hybrid quantum devices. \cite{xiang_2013,kurizki_2015} Strong and ultra strong coupling \cite{mills_1974,cao_2015,rameshti_2015} between magnons and microwave photons has been realized due to low damping and high spin density in YIG magnetization. \cite{huebl_2013,zhang_2014,tabuchi_2014,goryachev_2014,bai_2015} Due to possibility of coupling magnon modes to various oscillators, cavity photons are good candidates for mediating long distance indirect coupling of hybrid systems. Examples of different systems coupled using this approach are spin ensembles, \cite{schuster_2010,amsuss_2011} double quantum dots, \cite{frey_2012} and hybrid systems \cite{marcos_2010}.

Cavity mediated dispersive coupling between two magnetic systems has been discussed both theoretically \cite{rameshti_2018} and experimentally. \cite{lambert_2016} One of such systems has been proposed by Zhang \etal in Ref. \onlinecite{zhang_2015}, where they show that coherent superposition of coupled magnon states generates magnon "dark" and "bright" modes, formed due to out of phase and in phase oscillations in two magnons, respectively. The key property of "dark" mode is that it is decoupled from the cavity which enhances the coherence time, providing a platform to implement magnon gradient memory. \cite{zhang_2015} Existence of  "dark" modes has also been addressed in antiferromagnets. \cite{yuan_2017,xiao_2019} Realization of "dark" mode memory in Ref. \onlinecite{zhang_2015} is based on encoding information into the bright mode with subsequent conversion of the mode into "dark" with enhanced coherence time.

Due to inherent dissipative nature \cite{zhang_2017} of cavity and spin systems, the spin-photon coupling is not limited to coherent interactions. It was proposed recently, that dissipative spin-photon coupling \cite{grigoryan_2018,harder_2018} reveals mode level attraction at exceptional points (EP), which opens new avenue for exploring cavity-spintronics in the context of non-Hermitian physics. \cite{grigoryan_2019,bhoi_2019,cao_2019,zhang_2019,yang_2019} The nontrivial topology of the EP leads to coalescing of two eigenstates with phase difference of $\pm\pi/2.$ This leads to chirality of the eigenstate. \cite{heiss_2001,gao_2018} Together with exciting new effects in light-matter interactions, \cite{grigoryan_2018,bhoi_2019,cao_2019,zhang_2019,yang_2019} the discovery of dissipative spin-photon coupling reveals new opportunities of exploring hybridization of collective spin modes.


Here, we address the cavity mediated dispersive coupling between spatially separated magnetizations in the presence of phase-controllable fields on both FIs. First, we reproduce the results of dispersive spin-spin coupling in the absence of phase shifted field, where the "dark" and "bright" modes are obtained from the microwave signal transmission through the cavity. \cite{lambert_2016,rameshti_2018,xiao_2019} When both FIs are exposed to phase shifted field, we obtain mode level anticrossing with opposite order of "dark" and "bright" modes. When only one of the magnetizations is under the action of phase shifted field, the indirect spin-spin coupling becomes dissipative with mode level attraction. In contrast to coherent coupling, \cite{lambert_2016,rameshti_2018} where collective modes are formed from $ m_1 \pm m_2$ (depending on sign of the effective coupling, \cite{filipp_2011,rameshti_2018}) here we show that the chiral modes are formed as $m_1\pm i m_2,$ where the sign depends on which FI is under the phase shifted field. $m_{i}$ is the magnetization direction in $i$th FI. Moreover, we show that by either changing the phase shifted field or detuning between two ferromagnetic resonance (FMR) frequencies, we can change the chirality of the state.  
The model of dissipatively coupled oscillators in this approach can be applied in variety of alternative systems such as magnon-superconducting qubit coupling, \cite{quirion_2017}  hybridization between two mechanical modes. \cite{shkarin_2014}
\begin{figure}[t!]
	\includegraphics[width=\columnwidth]{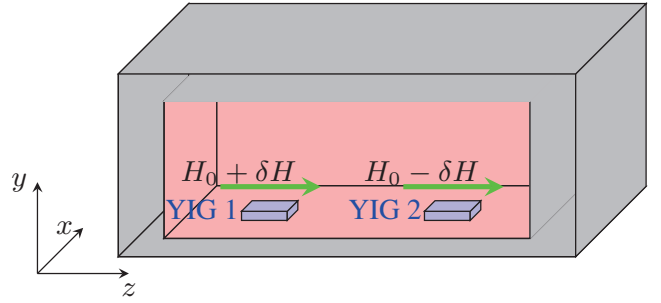}
	\caption{Schematic picture of the system}\label{fig:schem}
\end{figure} 
\section{Theoretical formalism}\label{sec:theory}
In \Figure{fig:schem} we schematically illustrate the system, where two magnetic materials are placed in a microwave cavity. We assume that the FIs are placed far from each other to ensure isolation and exclude direct coupling between the magnetizations. Our calculations are based on semiclassical model, where the microwave oscillations in the cavity is represented by an effective LCR circuit equation and Landau-Lifshitz-Gilbert (LLG) equation \cite{bloembergen_1954,bai_2015,grigoryan_2018} describes the dynamics of spin in magnetic materials. Faraday induction of FMR \cite{silva_1999} and the magnetic field created by Ampere's law \cite{bai_2015} are two classical coupling mechanisms. We assume that the crystal anisotropy, dipolar and external magnetic fields are in $\hzz$ direction.
The effective LCR circuit for the cavity is \cite{bloembergen_1954,bai_2015,grigoryan_2018,grigoryan_2019}
\begin{equation}
L \dot{\bj} +R \bj +\smlb{1/ C}\int \bj dt=\bV^F \label{eq:LCR},
\end{equation}
where $L,$ $C$, and $R$ represent the induction, capacitance, and resistance,  respectively. The current $\bj$ 
oscillates in $\hxx$-$\hyy$ plane. The driving voltage $\bV^F$ is induced from precessing magnetization of two FIs according to Faraday induction
\begin{figure*}[t!]
	\begin{tabular}{ccc}
		\includegraphics[width=.66\columnwidth]{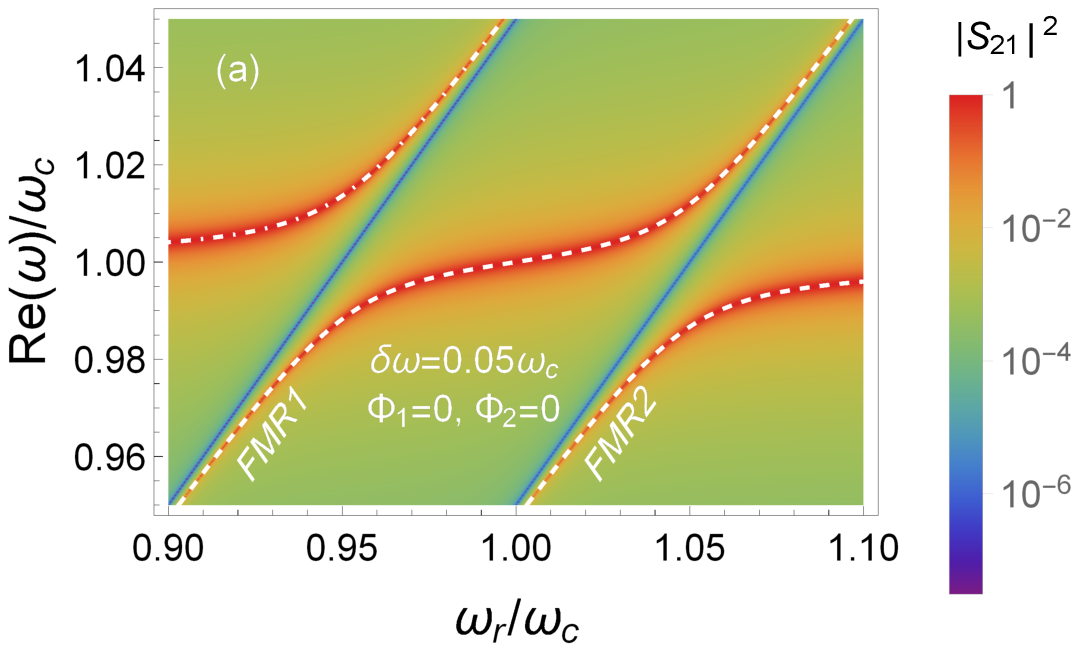}&		\includegraphics[width=.66\columnwidth]{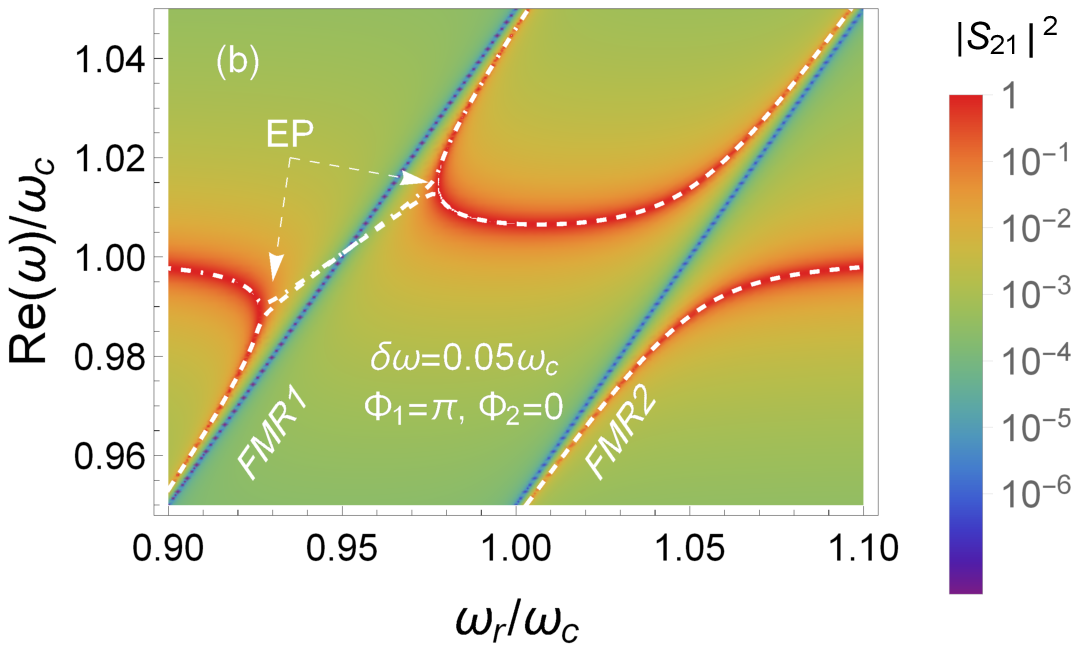}&
		\includegraphics[width=.66\columnwidth]{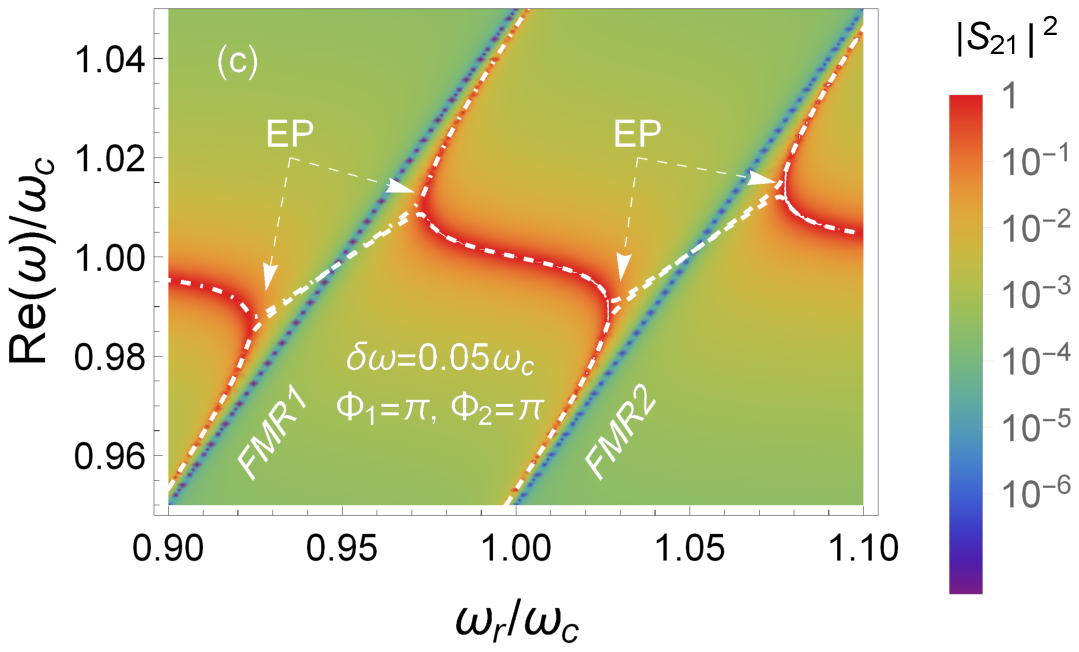}\\
				\includegraphics[width=.57\columnwidth]{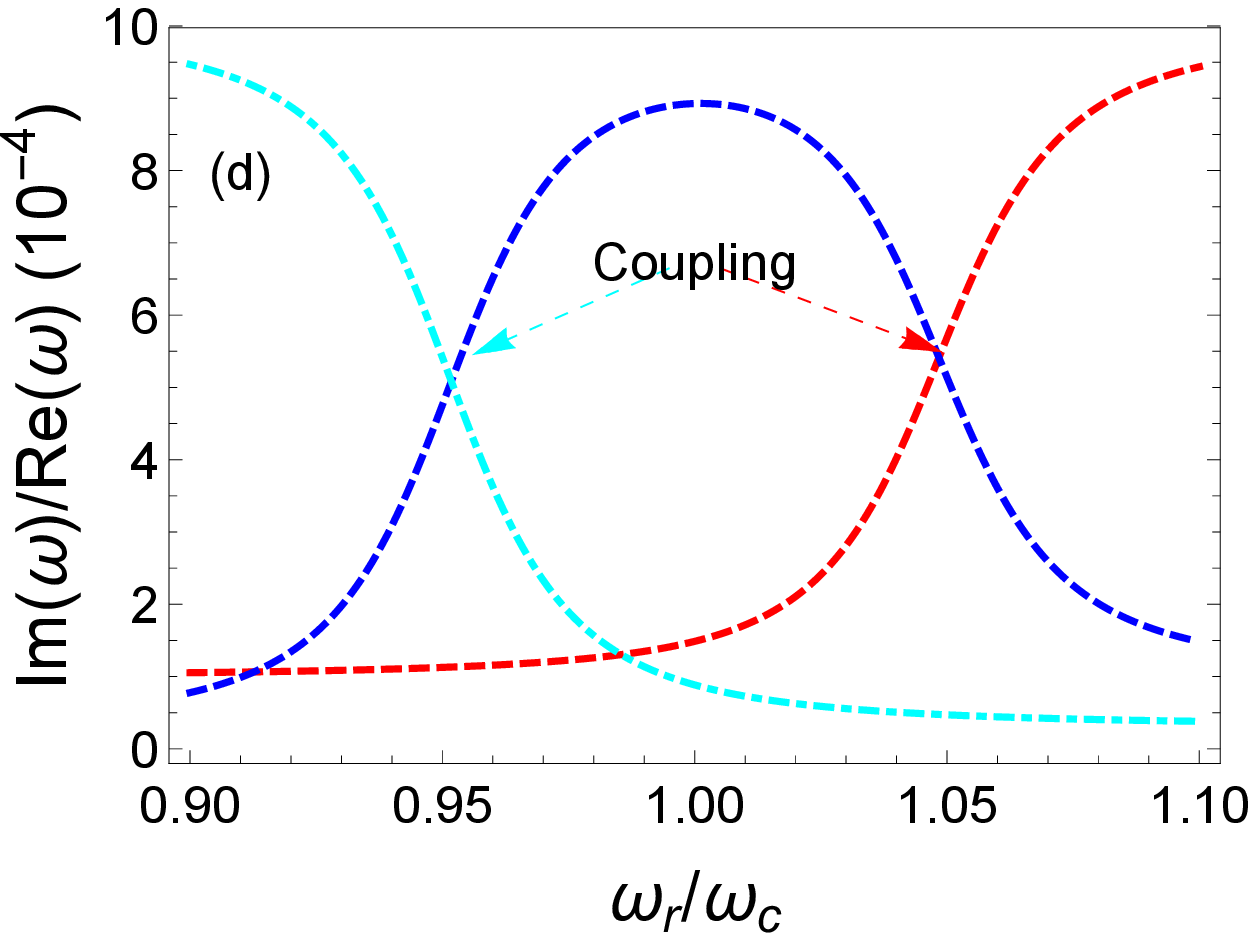}&
				\includegraphics[width=.6\columnwidth]{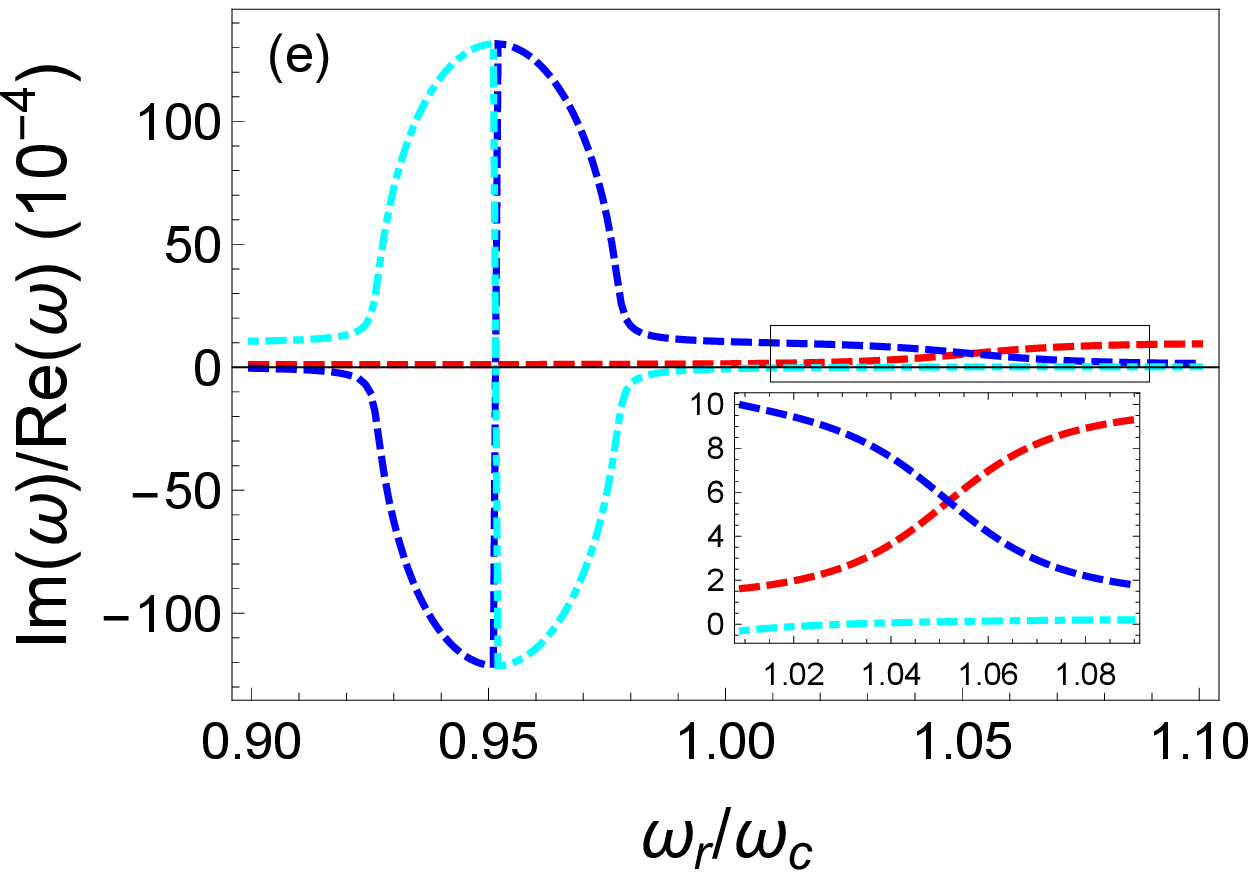}&
				\includegraphics[width=.6\columnwidth]{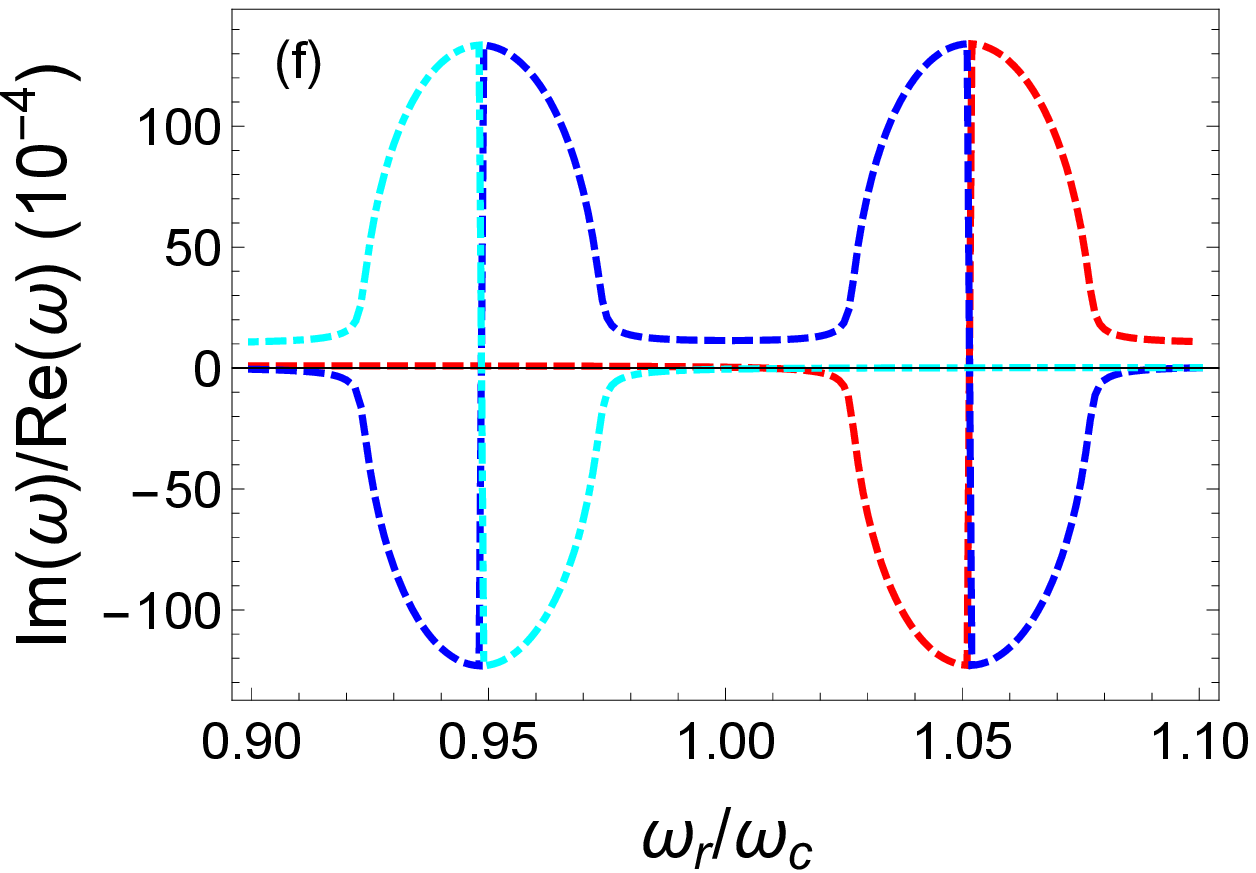}
	\end{tabular}\caption{First row shows the transmission amplitude in case of opposite detuning of the external magnetic field at each FI ($\omega_{r,1}/\omega_c=\omega_r/\omega_c+0.02,~ \omega_{r,2}/\omega_c=\omega_r/\omega_c-0.02$) for (a) $\Phi_1=\Phi_2=0,$ (b) $\Phi_1=\pi,$ $\Phi_2=0,$ and (c) $\Phi_1=\Phi_2=\pi.$ The dashed lines represent the normalized spectrum ($\re\omega/\omega_c$). (d), (e), and (f) in second row show the normalized damping of the system ($\im\omega/\re\omega$)  corresponding to parameters for (a), (b), and (c), respectively.}\label{fig:2}
\end{figure*} 
\begin{align}
&V^F_x\smlb{t}=\sum_j K_{c,j} L\dot{m}_{y,j},~V^F_y\smlb{t}=-\sum_j K_{c,j} L\dot{m}_{x,j}\label{eq:2},
\end{align}
where $j=1,2$ stands for first and second FI. $K_{c,j}$ is coupling parameter. The magnetization precession in the magnetic samples is governed by the LLG equation \cite{bloembergen_1954,bai_2015,grigoryan_2018}
\begin{align}
& \dot{\mb}_j=\gamma_j \mb_j\times \bH_j-\alpha_j \mb_j\times \dot{\mb}_j                 \label{eq:LLG},
\end{align}
\begin{figure*}[t!]
	\begin{tabular}{ccc}
		\includegraphics[width=.66\columnwidth]{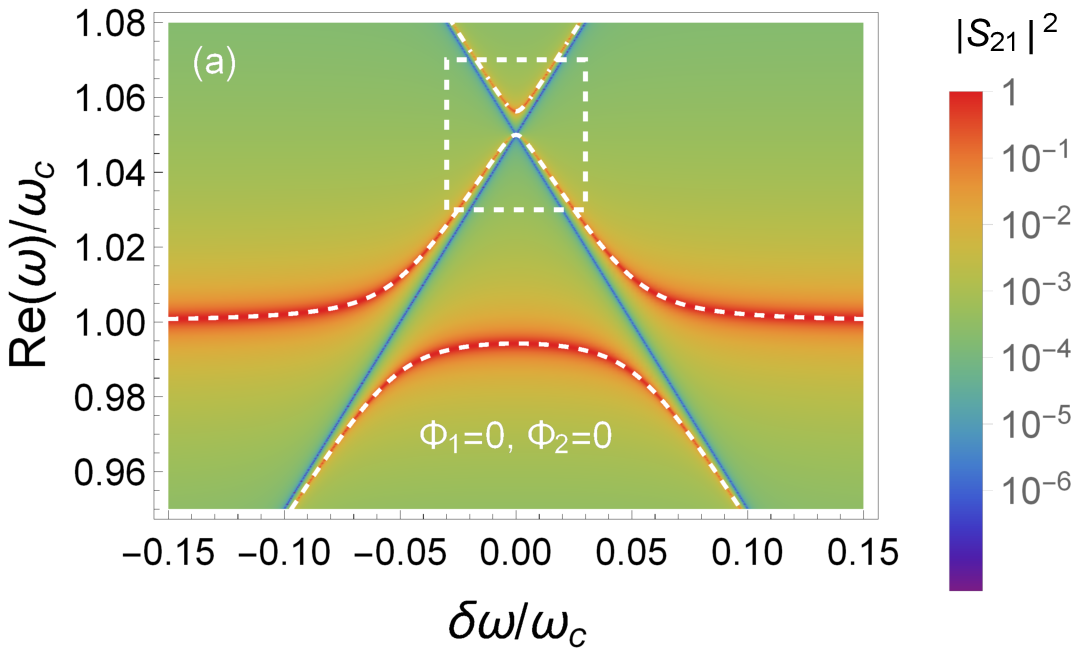}&		\includegraphics[width=.66\columnwidth]{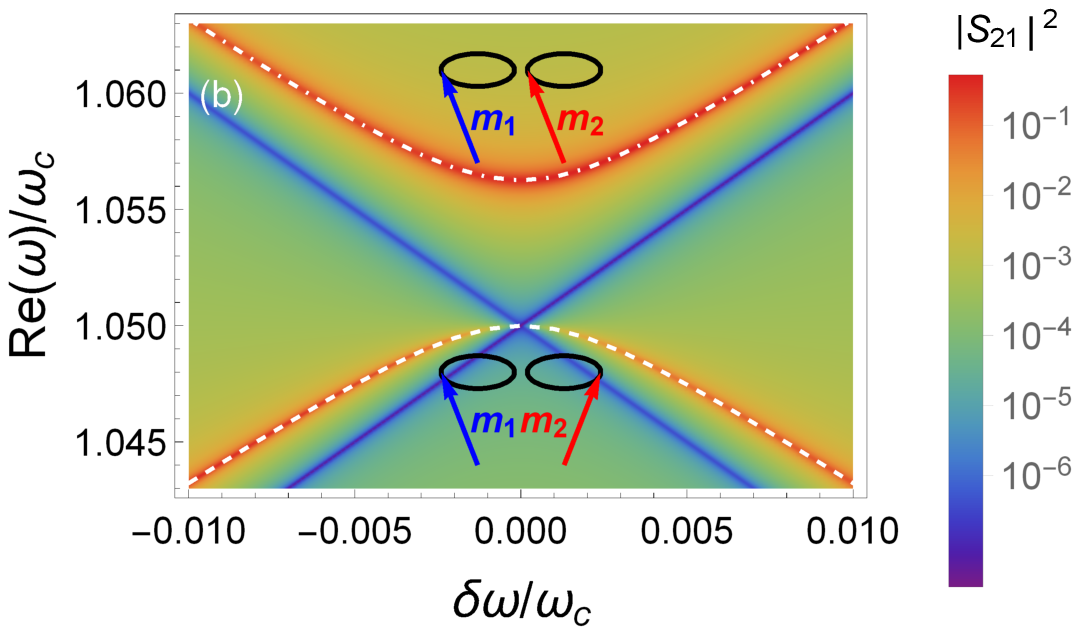}&
		\includegraphics[width=.53\columnwidth]{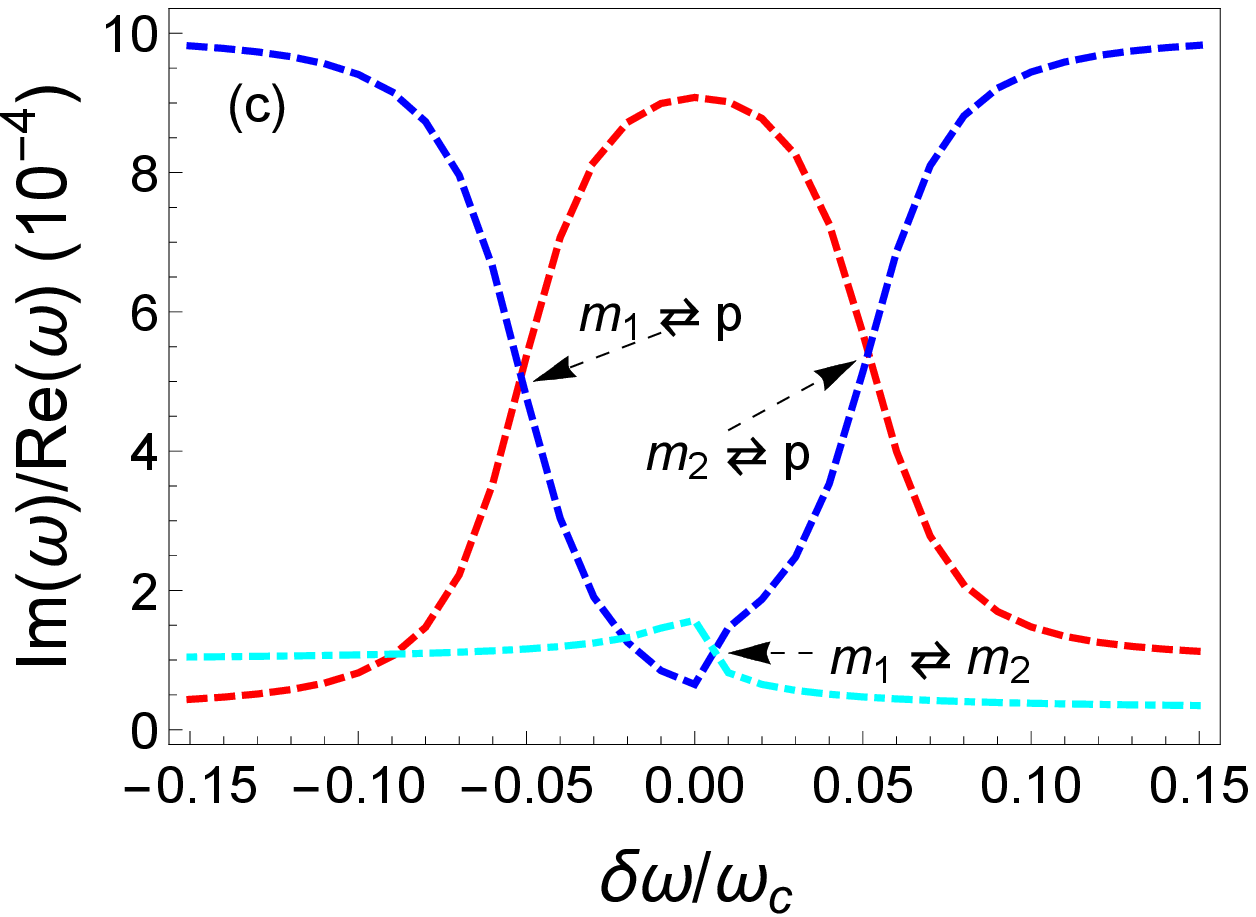}	\\	\includegraphics[width=.66\columnwidth]{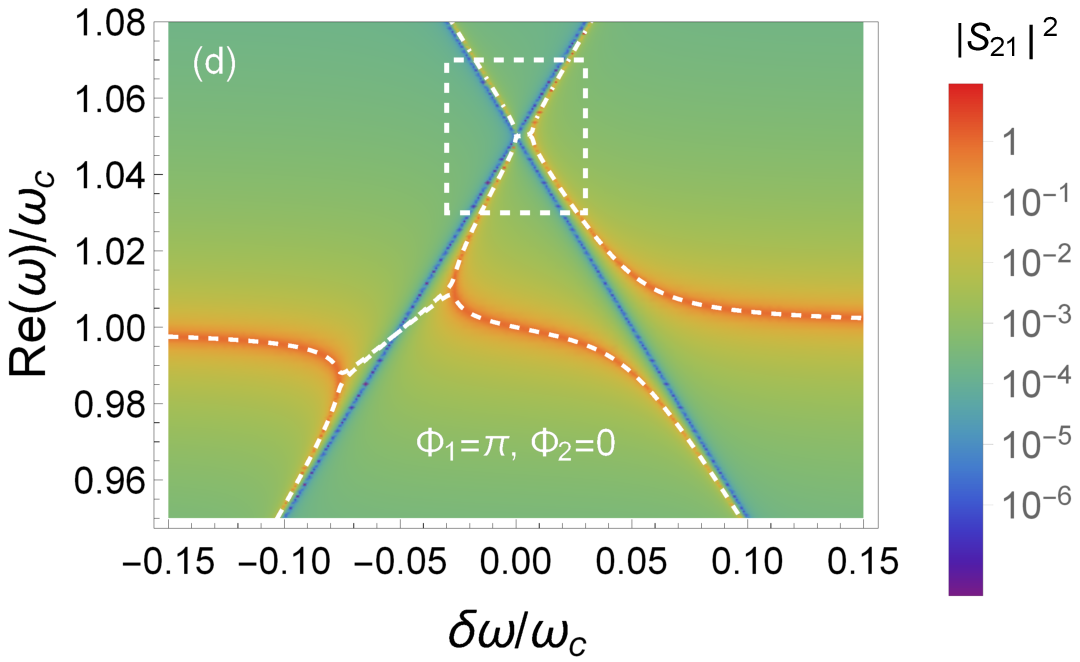}&		\includegraphics[width=.66\columnwidth]{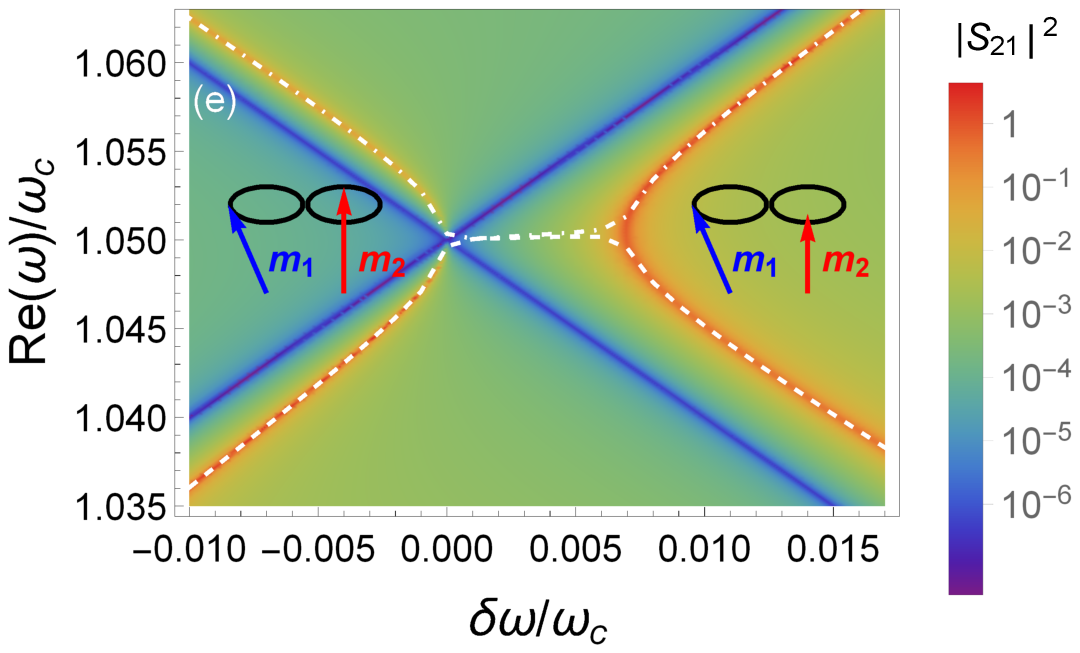}&
		\includegraphics[width=.55\columnwidth]{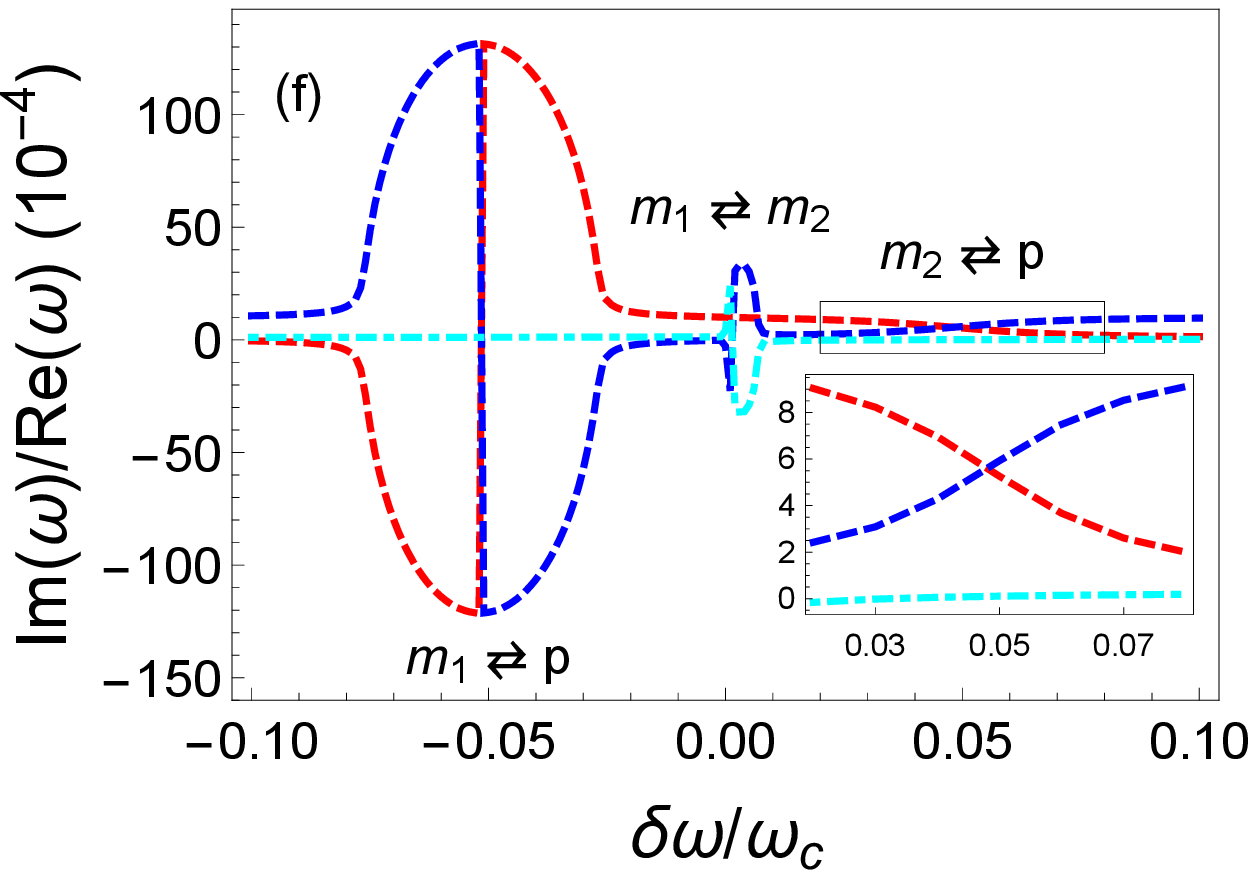}\\	\includegraphics[width=.66\columnwidth]{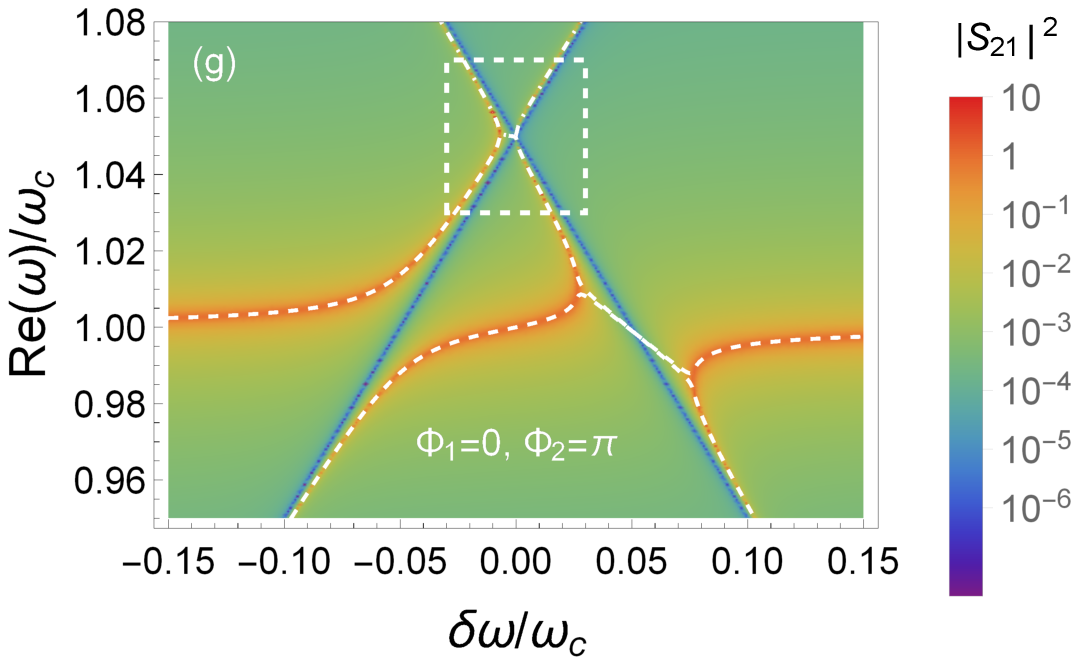}&		\includegraphics[width=.66\columnwidth]{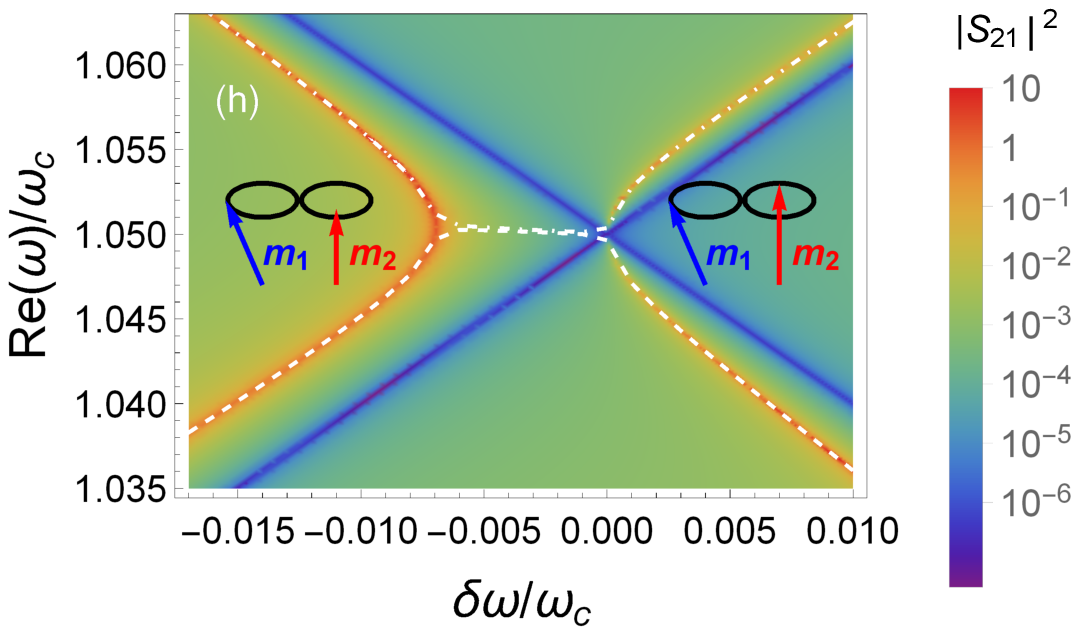}&
		\includegraphics[width=.56\columnwidth]{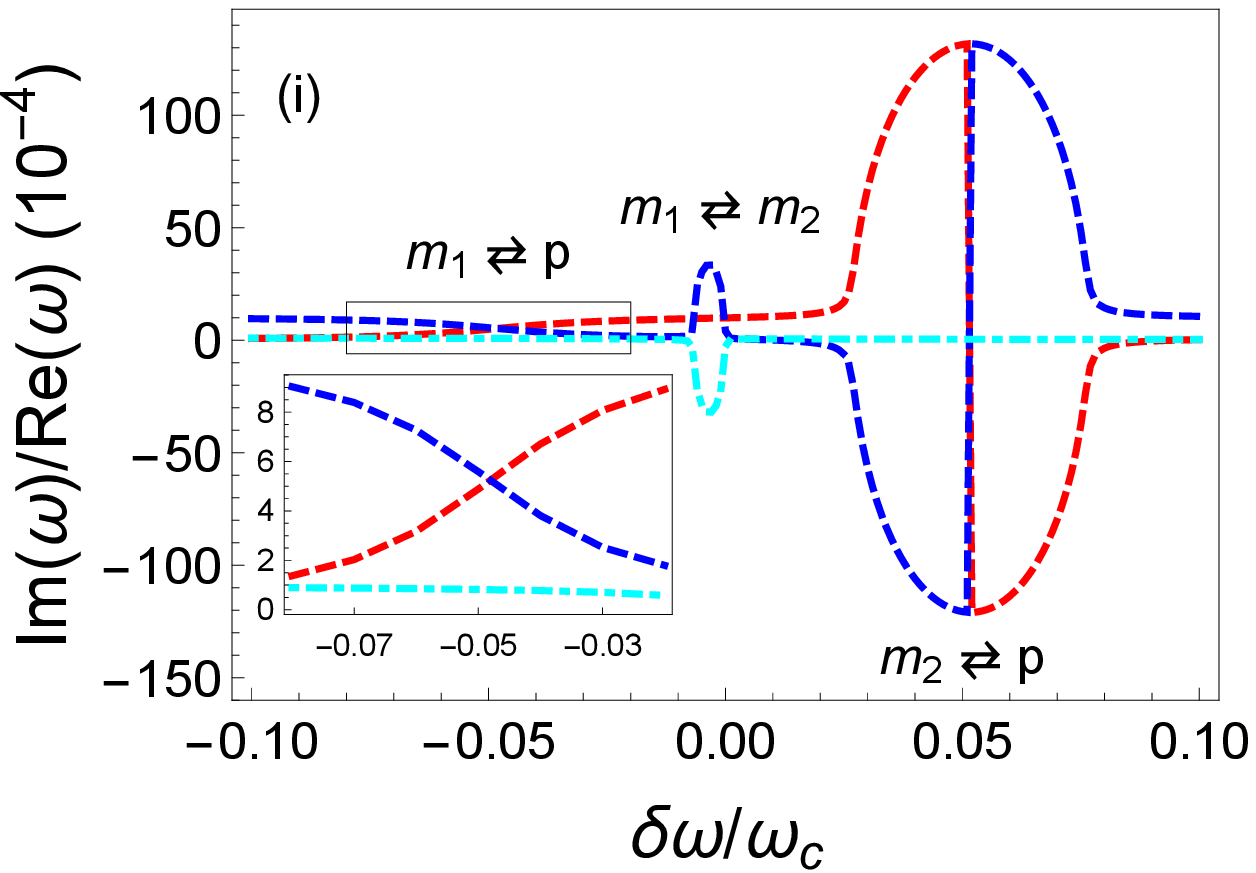}\\	\includegraphics[width=.66\columnwidth]{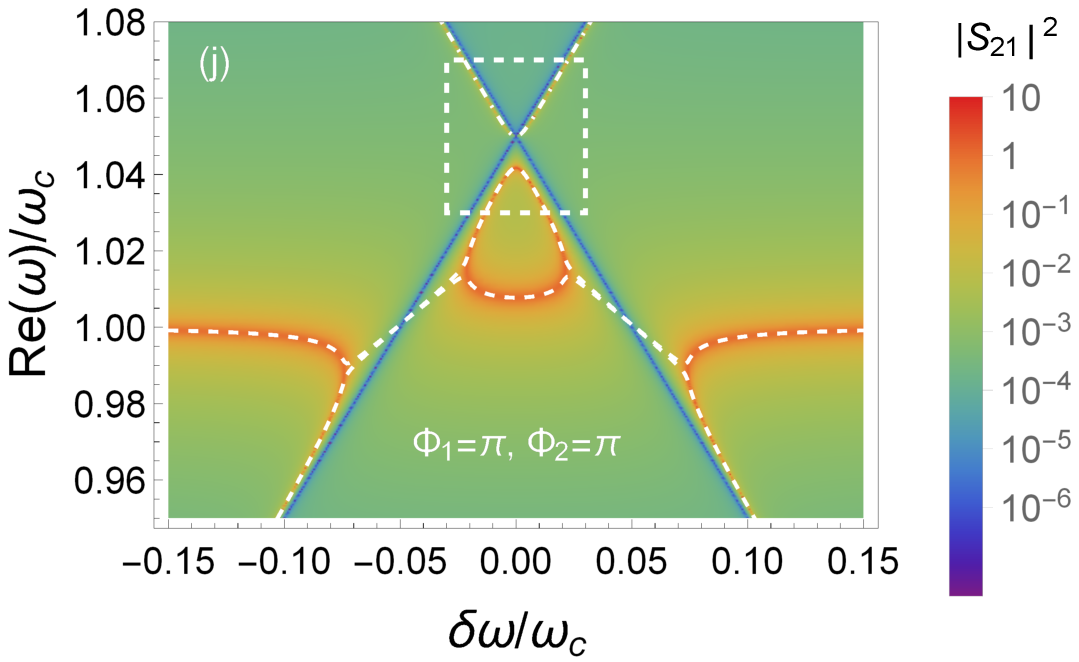}&		\includegraphics[width=.66\columnwidth]{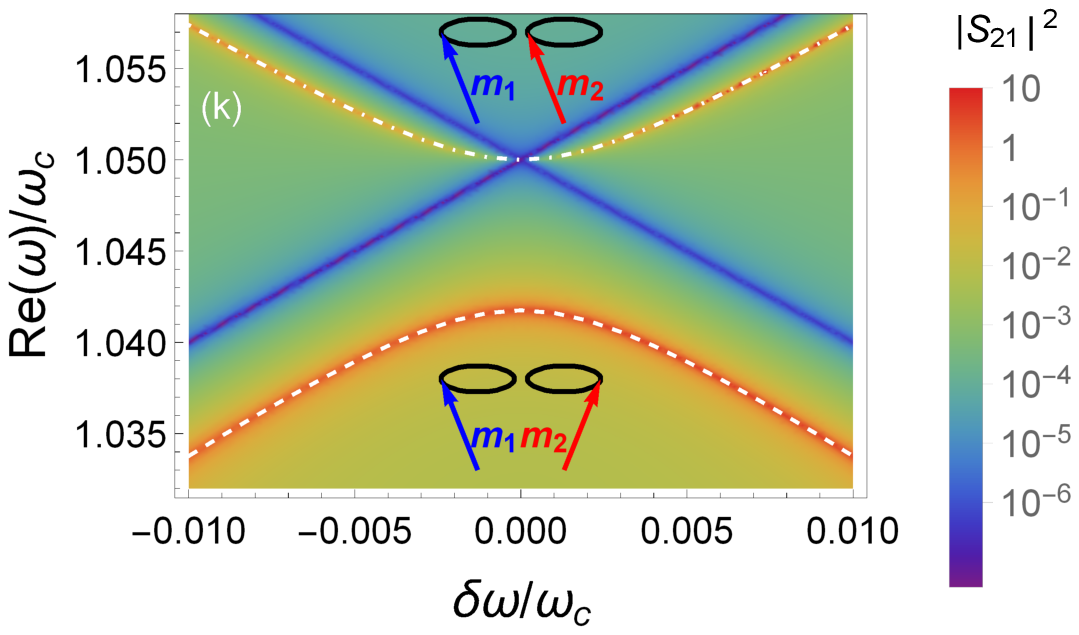}&
		\includegraphics[width=.53\columnwidth]{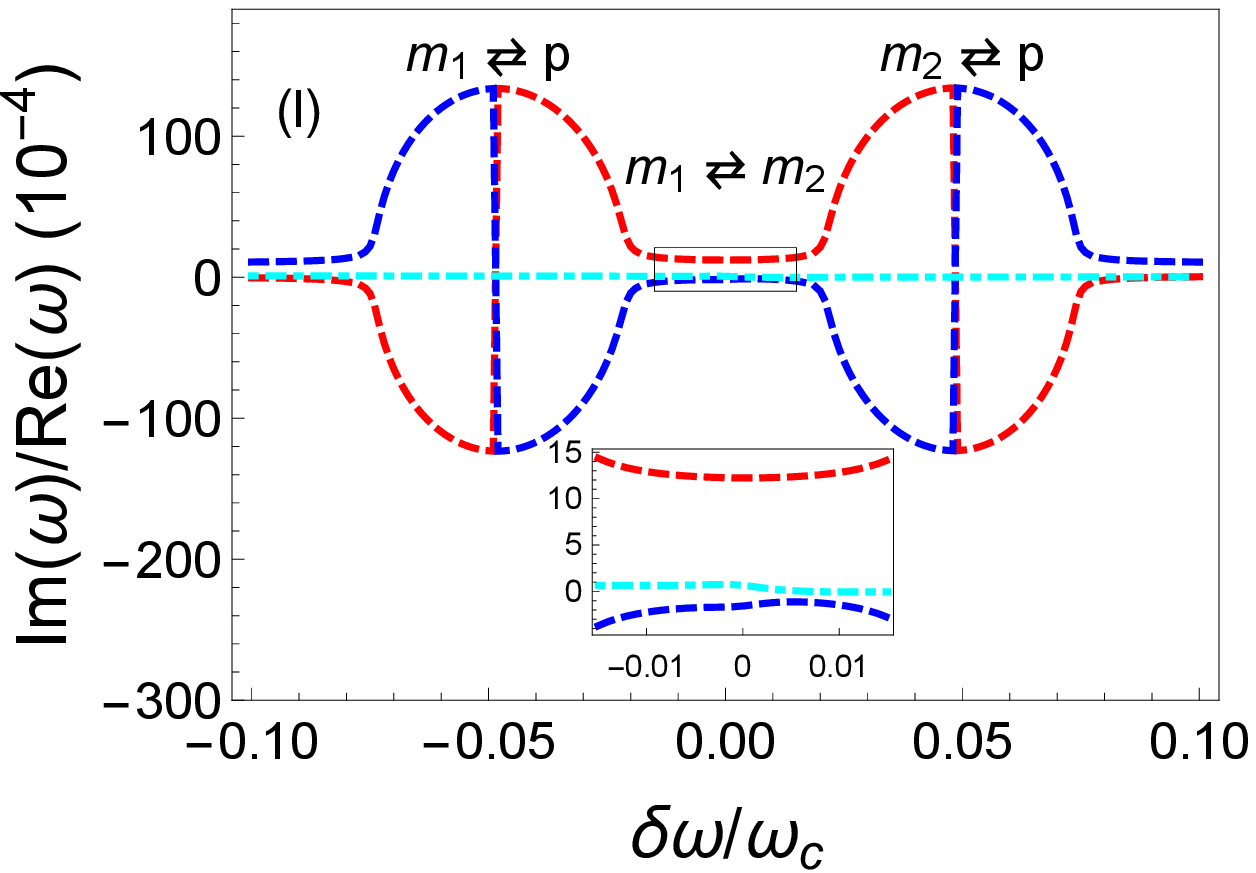}
	\end{tabular}\caption{First column shows the transmission amplitude dependence on the applied field detuning, where the dashed lines depict the normalized spectrum ($\re{\omega}/\omega_c$). The second column is the zoom of the white dotted boxes in corresponding plots in the first column. Third column shows the normalized damping ($\im{\omega}$/ $\re{\omega}$). The parameters are for first raw ((a), (b), (c)) $\Phi_1=\Phi_2=0,$ for second raw ((d), (e), (f)) $\Phi_1=\pi,$ $\Phi_2=0,$ for third raw ((g), (h), (i)) $\Phi_1=0,$ $\Phi_2=\pi,$ and for the last raw ((g), (h), (i)) $\Phi_1=\Phi_2=\pi.$ }\label{fig:transmission2}
\end{figure*}
where $\mb_j =\bM_j/M_{s,j},$ is the magnetization direction in $j$th FI. $M_{s,j},$ $\alpha_j,$ and $\gamma_j$ are the saturation magnetization, the intrinsic Gilbert damping parameter and gyromagnetic ratio, respectively. $\bH_j=\bH_{0,j}+e^{i\Phi_j} \bh^\ssf{A}$ is the effective magnetic field acting on the magnetization in $j$th FI, where $\bH_{0,j}=H_{0,j}\hzz$ is the sum of external, anisotropy and dipolar fields aligned with $\hzz$ direction. Based on our recent proposed mechanism of controlling phase $\Phi_j$ by introducing relative phase of microwaves in the cavity \cite{grigoryan_2018} and other mechanisms (including Lenz effect \cite{harder_2018} and inverted pattern of split ring resonator, \cite{bhoi_2019}) we assume $\Phi_j$ to be a free phase parameter \cite{grigoryan_2018} and $ \bh^\ssf{A}= \bh e^{-i\omega t}.$ Using $\mb=\hzz+m_\perp e^{-i\omega t}$ the LLG equation can be linearised  
\begin{align}
& m^+_j \smlb{\omega-\omega_{r,j}+i\alpha_j\omega}+e^{i\Phi_j}\omega_{m,j}h^+=0  \label{eq:linLLG},
\end{align}
where $m^+_j=m_{x,j}+im_{y,j}$ is the in-plane magnetization in $j$th FI, $\omega_{m,j}=\gamma_j M_{s,j}$, FMR frequency is $\omega_{r,j}\simeq \gamma_j H_{0,j}.$ The in-plane magnetic field is $h_+=h_x+ih_y.$ Using the form $\bj=\bj_\perp e^{-i\omega t}$ for solution of the LCR equation \Eq{eq:LCR} we obtain the system of coupled equations 
\begin{widetext}
\begin{equation}
 \Omega\smatrix{m_1^+\\m_2^+\\h^+}=0 \qwith \Omega=\smatrix{\omega+i\alpha_1\omega-\omega_{r,1}&0&e^{i\Phi_1} \omega_{m,1}\\0&\omega+i\alpha_2\omega-\omega_{r,2}&e^{i\Phi_2} \omega_{m,2}\\\omega^2 K_1^2 &\omega^2 K_2^2&\omega^2+2i\beta \omega \omega_c-\omega_c^2}\label{eq:matr1}
\end{equation},
\end{widetext}
where from Ampere's law we have the magnetic field of the microwave, which exerts torque on the FI magnetization
\begin{equation}
h_x=K_m j_y,~h_y=-K_mj_x \label{eq:ampere},
\end{equation}
with $K_m$ being the coupling parameter and $K_j\simeq \sqrt{K_{c,j} K_m}.$ The cavity frequency is $\omega_c=1/\sqrt{LC}$ and $\beta=R/\smlb{2L\omega_c}$ stands for the cavity mode damping. From solution of $\det\Omega=0$ in \Eq{eq:matr1} we use three positive roots of $\omega.$ The real and imaginary components of $\omega$ determine the spectrum and damping of the system, respectively.
\section{Results and Discussion}\label{sec:results}
We calculate the transmission amplitude using input-output formalism \cite{bai_2015,grigoryan_2018
}
\begin{align}
&\Omega  \smatrix{m_1^+\\m_2^+\\h^+}=\smatrix{0\\0\\ \omega ^2 h_{0}} , ~S_{21}=\Gamma h^+ / h_{0}
\label{eq:transmission},
\end{align}
where $h_{0}$ is the input magnetic field driving the system, $\Gamma$ is a normalization parameter. \cite{bai_2015,grigoryan_2018} We first discuss the case, where magnetic fields on two FIs are detuned with opposite signs ($H_{0,1\smlb{2}}=H_0 \pm \delta H$) $\omega_{r,1,2}=\omega_{r}\pm \delta \omega,$ where $\delta\omega=0.05\omega_c.$ We use different Gilbert dampings for FIs, $\alpha_1=3\times 10^{-5},$  $\alpha_2= 10^{-4},$ which are relevant with experimental values. \cite{harder_2018} The cavity mode frequency is $\omega_c/2\pi=13.2$ GHz with cavity damping $\beta= 10^{-3},$ $\omega_{m,1}=\omega_{m,2}=\gamma  M_s =0.36 \omega_c,$ where $\mu_0 M_s=0.178$T and $\gamma/2\pi= 27 \mu_0 $GHz/T. \cite{harder_2018} Coupling constant is $K_1=K_2=0.03.$ The coloured area in the first row of \Figure{fig:2} is the transmission amplitude for different values of $\Phi_j$ as a function of frequency $\re\omega$ (normalized by $\omega_c$) and uniform magnetic field $\omega_{r}$. The dashed lines show the spectrum $\re{\omega(\omega_{r})}$. Corresponding linewidth evolutions are shown in the second row of \Figure{fig:2}. For $\Phi_1=\Phi_2=0$ we reproduce two distinct anticrossings in \Figure{fig:2} (a) with two characteristic peaks of transmission indicating coupling of two magnetizations with the cavity mode. \cite{rameshti_2018,lambert_2016} Linewidth exchange \cite{harder_2018} between cavity mode with FI modes at resonant frequencies is shown in \Figure{fig:2} (d). 

In \Figure{fig:2} (b) we show the transmission amplitude and corresponding spectrum for $\Phi_1=\pi$ and $\Phi_2=0.$ It is seen that while $\Phi_2=0$ leads to usual coupling with transmission peaks at anticrossing near $\omega_{r}=1.05 \omega_c,$ the phase parameter ($\Phi_1=\pi$) from first FI causes mode level attraction \cite{grigoryan_2018,harder_2018,bhoi_2019} and coalescence of the modes at two EPs. Corresponding repulsion of linewidth \cite{harder_2018,bhoi_2019} for $\Phi_1=\pi$ is shown in \Figure{fig:2} (e), where the inset shows evolution of the linewidth for second FI, where the phase parameter is $0$. In \Figure{fig:2} (c) and (f) we plot the spectra of real and imaginary components of $\omega$ for $\Phi_1=\Phi_2=\pi$, respectively. Attraction of real and repulsion of imaginary components of $\omega$ is seen at resonant magnetic field of both FIs.

After discussing the resonant coherent and dissipative coupling between two FIs and the cavity we move into the dispersive regime where the FMR frequencies of FIs are significantly detuned from cavity mode $\abs{\Delta}\equiv \abs{\omega_{r,1,2}-\omega_c}> K_{1,2}\omega_{m,1,2}.$ We do so by adjusting the magnetic field on FIs ($\omega_{r}=1.05\omega_c$) and study effect of detunings $\delta \omega$ (normalized by $\omega_c$) in the dispersive regime. In \Figure{fig:transmission2} (a) we plot the transmission as a function of $\omega$ and $\delta \omega$ for $\Phi_1=\Phi_2=0,$ meaning that there is no phase shift introduced in either coupled system. It is seen that the coupling anticrossings between FMR modes and cavity mode appear at larger detuning, when the effective FMR frequencies are in resonance with the cavity mode. More interestingly, an anticrossing between two FMR modes appears at $\delta\omega=0,$ which indicates cavity-mediated coupling between two FIs. \cite{lambert_2016,rameshti_2018} The boxed part of the plot is zoomed in \Figure{fig:transmission2} (b), where we can see the characteristic anticrossing of two Kittel modes of two FIs. \cite{lambert_2016,rameshti_2018} We can also observe the "dark" and "bright" modes, where the latter has larger oscillator strength than the former one \cite{filipp_2011,lambert_2016,rameshti_2018}. In \Figure{fig:transmission2} (c) we plot the imaginary components of $\omega$, that is linewidth of the system. Characteristic linewidth exchange \cite{bai_2015,harder_2018} between cavity mode and FMR modes is seen for large detuning ($\delta\omega=\pm0.05\omega_c$). Similarly, linewidth exchange between two FMR modes occurs at $\delta\omega=0,$ indicating coherent coupling between two FIs.

Next, we set one of the phase parameters to be $\Phi_1=\pi$ while keeping $\Phi_2=0.$ This corresponds to a situation, when the second FI is coherently coupled with the cavity while first one is in dissipative coupling regime. \cite{grigoryan_2018,harder_2018} The transmission and spectrum for this set of parameters is shown in \Figure{fig:transmission2} (d). According to the phase parameter, the spectrum in first FI-cavity coupling region ($\delta\omega=-0.05\omega_c$) shows level attraction, while level repulsion occurs at second FI-cavity coupling region ($\delta\omega=0.05\omega_c$). As it is seen from boxed area of \Figure{fig:transmission2} (d) and zoomed in \Figure{fig:transmission2} (e) the spectrum of two coupled FIs also shows level attraction feature, indicating dissipative spin-spin coupling. An interesting feature of in the transmission amplitude at this region is that the "dark" and "bright" modes are formed as a collective mode with phase difference equal to $\pm \pi/2,$ which will be discussed in details later. \Figure{fig:transmission2} (f) shows the corresponding damping dependences on the detuning $\delta\omega.$ Inset shows typical damping exchange for FI-2 at positive detuning ($\delta\omega/\omega_c =0.05$) as that in \Figure{fig:transmission2} (c). Beside the large linewidth repulsion for $\delta\omega=-0.05\omega_c,$ corresponding to dissipative coupling between the magnetization in FI-1 and cavity photons, similar feature is seen at $\delta\omega=0$ for dissipative spin-spin coupling. In \Figure{fig:transmission2} (g-i) we show the same as in (d-f) for $\Phi_1=0,$ $\Phi_2=\pi.$ One can see in \Figure{fig:transmission2} (g) and zoomed in (h) that the order of "dark" and "bright" modes is shifted compared to $\Phi_1=\pi,$ $\Phi_2=0$ case.

In \Figure{fig:transmission2} (j) (zoomed picture of the boxed part in (k))  we plot the transmission and the spectrum when both phase parameters are $\Phi_1=\Phi_2=\pi.$ For large negative/positive values of the detuning ($\delta\omega =\pm 0.05\omega_c$) both FIs' magnetizations are dissipatively coupled with the cavity modes. Corresponding linewidth repulsion is shown in \Figure{fig:transmission2} (l). It is seen in \Figure{fig:transmission2} (j) that, although both FIs are dissipatively coupled with the cavity mode, the spectrum of cavity-mediated coupling of FIs' magnetizations shows anticrossing feature. Correspondingly, as seen from the inset in \Figure{fig:transmission2} (k), the linewidth at $\delta\omega=0$ show exchange feature in contrast to linewidth repulsion at  $\delta\omega=\pm 0.05\omega_c.$ 


To better understand the spectrum and collective states of cavity-mediated dissipative magnon-magnon coupling, here we develop a quantum picture by considering the Hamiltonian ($\hbar=1$)
\begin{align}
&H=H_0+H_g, \qwith \nn
&H_0=\omega_c a^\dagger a+\sum_j \omega_{r,j} m_j^\dagger m_j, \nn
&H_g=\sum_j g_je^{i\Phi_j/2}\smlb{a^\dagger m_j+ m_j^\dagger a} \label{eq:hamquant},
\end{align} 
where the first and second terms in $H_0$ stand for cavity photon and $j$th ($j=1,2$) FI magnon energy, respectively. $\omega_c$ is the cavity mode frequency, $\omega_{r,1,2}=\omega_{r}\pm \delta \omega$ is the FMR frequency. $H_g$ is the coupling between them. $a\smlb{a^\dagger}$ and $m_j(m_j^\dagger)$ are annihilation (creation) operators for cavity photons and magnons in $j$th FI, respectively. $g_j$ is the coupling of $j$th magnetization with cavity and $\Phi_j$ is the phase parameter with $\Phi_j=0$ for coherent coupling \cite{blais_2004} and $\Phi_j=\pi$ for dissipative coupling. \cite{bernier_2018}) Next, we use Schrieffer-Wolff transformation \cite{schrieffer_1966,grigoryan_2013} 
\begin{equation}
H^\prime=e^\Lambda H e^{-\Lambda}=H+\midb{\Lambda,H}+\half \midb{\Lambda,\midb{\Lambda}}+\cdots, \label{eq:SW} 
\end{equation}
where by choosing a transformation operator $\Lambda$ such that 
\begin{equation}
H_g+\midb{\Lambda,H_0}=0\label{eq:SWoper}.
\end{equation} 
we eliminate the direct magnon-photon interaction in favour of higher order (up to second order of $g_j$) coupling  between magnetic moments  \cite{blais_2007,filipp_2011} in dispersive regime ($\abs{\Delta_j}\equiv \abs{\omega_{r,j}-\omega_c}>g_j$). The transformation operator satisfying condition \Eq{eq:SWoper} is $\Lambda= \sum_j g_{j} e^{i\Phi_j/2}\smlb{  m_j^\dag a-a^\dag m_j }/\Delta_j.$ From \Eq{eq:SW} we obtain 
\begin{align}
&H^\prime=H_{c}+H_{M}, \qwith \nn
&H_{c}=\omega_c^\prime a^\dag a ,\nn
&H_M=\sum_j \omega_{m,j}^\prime m_j^\dag m_j + g_{eff}\smlb{   m_1^\dag  m_2+   m_2^\dag  m_1} \label{eq:effham},
\end{align}
where $H_c$ is the cavity energy with $\omega_c^\prime = \omega_c- \sum_{j} e^{i\Phi_j} g_{j}^2/ \Delta_j$ is dispersive shift of the cavity frequency. $H_M$ in \Eq{eq:effham} being the magnetic Hamiltonian without coupling with cavity, where \begin{equation}
\omega_{r,j}^\prime =\omega_{r,j}+e^{i\Phi_j} {g_{j}^2 \ov \Delta_j} \label{eq:om}
\end{equation} is the the Lamb shift of the FMR frequency due to the presence of virtual photons. \cite{filipp_2011} Effective coupling between two FIs becomes\cite{blais_2007,filipp_2011} 
\begin{equation}
g_{eff}=  \half e^{i{\smlb{\Phi_1+\Phi_2}\ov 2}} g_{1} g_{2}\smlb{ {1\ov \Delta_1} +{1 \ov \Delta_2} }. \label{eq:geff}
\end{equation}  
For simplicity, we consider the case when $\Delta_1=\Delta_2\equiv \Delta$ and $g_1=g_2\equiv g.$ The eigenvalues of $H_M$ become
\begin{align}
&E^\pm=\half\smlb{\omega_{r,1}^\prime+\omega_{r,2}^\prime \pm \sqrt{\omega_g}},\qwith \nn
&\omega_g=\smlb{\omega_{r,1}^\prime-\omega_{r,2}^\prime}^2+4 g_{eff}^2 \label{eq:eigval}
\end{align}
Here we discuss four cases: (i) $\Phi_1=\Phi_2=0,$ (ii) $\Phi_1=\Phi_2=\pi,$ (iii) $\Phi_{1}=\pi,$ $\Phi_{2}=0,$ and (iv)  $\Phi_{1}=0,$ $\Phi_{2}=\pi.$ It follows from \Eqs{eq:geff}{eq:eigval} that in two former cases ($\Phi_1=\Phi_2=0(\pi)$) $\omega_g>0.$ At $\delta\omega=0,$ the higher and lower eigenstates of Hamiltonian in \Eq{eq:effham} can be written in general form as  
\begin{align}
& \Psi^\pm={1\ov \sqrt{2}}\smlb{m_1\pm e^{i{\smlb{\Phi_1+\Phi_2}\ov 2}} \text{sgn}\smlb{g^2\ov \Delta} m_2}  \label{eq:BD},
\end{align}
where $\Psi^+$ and $\Psi^-$ correspond to higher and lower energy states. In the absence of phase shift $\Phi_{1,2}=0,$ $\Psi^+$ and $\Psi^-$ correspond do "bright" and "dark" modes, respectively when $\text{sgn}\smlb{g^2/ \Delta}>0$. \cite{zhang_2015,filipp_2011,rameshti_2018}  Construction of "dark" mode memory proposed in Ref. \onlinecite{zhang_2015} is based on fast (faster than magnon dissipation rate) conversion between the "bright" and "dark" modes. For $\Phi_{1,2}=0,$ \Eq{eq:BD} reduces to coherent coupling discussed in Ref. \onlinecite{zhang_2015}. In this case, the conversion between "dark" and "bright" states can be realized by rapidly tuning the magnetic bias field, \cite{filipp_2011,rameshti_2018,zhang_2015} which is prohibited in the experiment due to slow response of the local inductive coils. \cite{zhang_2015} It follows from \Eq{eq:BD} that in our proposal, the conversion can be realized by tuning the phase parameters $\Phi_{1,2}.$ The parameters can be tuned by additional microwave applied to FIs \cite{grigoryan_2018,bhoi_2019} and thus, does not suffer from the slow response of magnetic field.
For positive sign of $g^2/\Delta$ the "bright" ($B$) and "dark" ($D$) eigenstates become for (i)
\begin{align}
&B^{\smlb{\text{i}}}=\smlb{m_1+m_2}/\sqrt{2},& D^{\smlb{\text{i}}}=\smlb{m_1-m_2}/\sqrt{2},\label{eq:BD1},
\end{align}
and for (ii)
\begin{align}
&B^{\smlb{\text{ii}}}=\smlb{m_1-m_2}/\sqrt{2},& D^{\smlb{\text{ii}}}=\smlb{m_1+m_2}/\sqrt{2}\label{eq:BD1.1},
\end{align}
Opposite order of "dark" and "bright" collective modes is shown in \Figure{fig:transmission2} (b) and (k), where the former one corresponds to (i) and the latter one is for (ii).

\begin{figure}[t!]
	\includegraphics[width=\columnwidth]{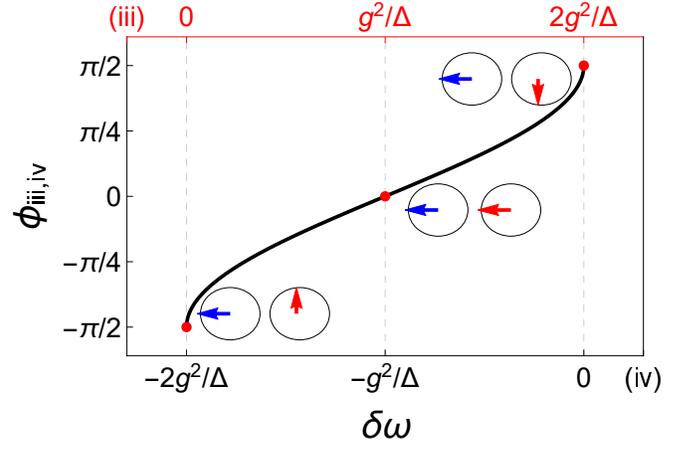}
	\caption{Dependence of phase shift on detuning. Upper frame ranges from $0$ (EP1) to $2g^2/\Delta$ (EP2) for (iii) and lower frame ranges from $-2g^2/\Delta$ (EP1) to $0$ (EP2) for (iv). Arrows in cartoon demonstrate the phase difference between two magnetization in collective mode. Red dotes correspond to values for the cartoons.}\label{fig:phase}
\end{figure} 

We now move to discussion of the cavity-mediated coupling between two FIs when one of the magnetization is coupled to the cavity dissipatively, while other is coherently coupled, corresponding to (iii) and (iv). From \Eq{eq:geff} the effective coupling ($g_{eff}$) in this case becomes imaginary, which, in analogy with dissipative coupling in \Eq{eq:hamquant}, leads to level attraction between two FMR modes and coalescence at EPs. This feature is shown \Figure{fig:transmission2} (e) for (iii) and (h) for (iv). Coalesced two energy levels at EPs lead to coalescing of two eigenstates at EPs and a single eigenvector with a single eigenvalue survives. \cite{heiss_2012,grigoryan_2018,grigoryan_2019,bernier_2018} It follows from \Eq{eq:eigval} that the band closing at EPs occurs when $\omega_g=0$. Taking into account the Lamb shift of the FMR frequencies (\Eq{eq:om}), positions of the EPs for (iii) are $\delta\omega=0$ and $\delta\omega=2g^2/\Delta$ (\Figure{fig:transmission2} (e)). Similarly, the EPs for (iv) are at $\delta\omega=-2g^2/\Delta$ and $\delta\omega=0$ (see \Figure{fig:transmission2} (h)). The eigenstates at range of coupling bandwidth (frequencies between two EPs) for (iii) and (iv) are calculated to be 
\begin{align}
&\Psi^{(\text{iii})}=\smlb{m_1+e^{i\phi_{\smlb{\text{iii}}}}m_2}/\sqrt{2},\nn
&\Psi^{(\text{iv})}=\smlb{m_1+e^{i\phi_{\smlb{\text{iv}}}}m_2}/\sqrt{2} \label{eq:iv}
\end{align}
where $\phi_{\smlb{\text{iii},\text{iv}}}$ is the phase leg between two modes. Dependence of the phase difference between two modes in collective mode is shown in \Figure{fig:phase}. It is seen that by tuning the detuning $\delta\omega$ from $0$ ($-2g^2/\Delta$) to $2g^2/\Delta$ ($0$) for (iii) ((iv)), we can shift the chirality of the state. Moreover, the same point $\delta\omega=0$ has opposite chirality for (iii) and (iv). The eigenstates at the EPs are calculated to be
\begin{align}
& B^{\smlb{\text{iii}}}=\smlb{m_1-i m_2}/\sqrt{2},
&D^{\smlb{\text{iii}}}=\smlb{m_1+i m_2}/\sqrt{2}, \nn
&B^{\smlb{\text{iv}}}=\smlb{m_1+i m_2}/\sqrt{2},
&D^{\smlb{\text{iv}}}=\smlb{m_1-i m_2}/\sqrt{2} . \label{eq:BD2}
\end{align}
It follows from \Eqs{eq:BD1}{eq:BD2} that fast switching of $\Phi_i$ allows to construct "dark" mode memory based on switching between collective modes with phase difference $0$ to $\pi,$ as well as between $\pi/2$ to $-\pi/2$.

In summary, we study dispersive coupling between magnetizations of two FIs mediated by dissipative spin-photon coupling. We show that when only one of the spin modes is dissipatively coupled to the cavity mode, the cavity mediated spin-spin coupling becomes dissipative, where the energy levels of two spin modes attract to each other. Varying the phase parameters in both FIs allows to construct "bright" and "dark" modes with tunable phase shift between two spin modes. Chiral modes with controllable chirality can be constructed when only one of FIs is under the action of phase shifted field. 



\begin{acknowledgements}
This work was financially supported by National Key Research and Development Program of China (Grant No. 2017YFA0303300) and the National Natural Science Foundation of China (No.61774017, No. 11734004, and No. 21421003).
\end{acknowledgements}
%


\end{document}